\documentclass{iau} 
\usepackage{graphicx}
\usepackage{hyperref}
\usepackage{natbib}
\usepackage{longtable}

\newcommand{\Ms}{\ensuremath{M_{\odot}}}
\newcommand{\cf}{{\it c.f.~}}
\newcommand{\ie}{{\it i.e.}}

\newcommand{\beq}{\begin{equation}}
\newcommand{\eeq}{\end{equation}}

\newcommand{\mphot}{\ensuremath{M_{\rm phot}}}

\newcommand{\tcr}{\ensuremath{\tau_{cr}}}
\newcommand{\mcl}{\ensuremath{M_{cl}}}
\newcommand{\rh}{\ensuremath{r_h}}

\newcommand{\reff}{\ensuremath{r_{\rm eff}}}

\newcommand{\logten}{\ensuremath{\log_{10}}}
\newcommand{\kmps}{\ensuremath{\rm~km~s}^{-1}}
\newcommand{\tauh}{\ensuremath{\tau_h}}

\newcommand{\mnras}{{\it MNRAS}}
\newcommand{\aap}{{\it A\&A}}
\newcommand{\apj}{{\it ApJ}}

\newcommand{\araa}{{\it ARA\&A}}

\title[Initial conditions for starburst clusters] 
{Initial conditions of formation of starburst clusters: constraints from stellar dynamics}

\author[S. Banerjee]   
{Sambaran Banerjee$^{1,2}$}

\affiliation{$^1$Argelander-Institut f\"ur Astronomie \\ 
Auf dem H\"ugel 71, D-53121 Bonn, Germany \\ Email: {\tt sambaran@astro.uni-bonn.de} \\[\affilskip]
$^2$Helmholtz-Instituts f\"ur Strahlen- und Kernphysik\\
Nussallee 14-16, D-53115 Bonn, Germany
}

\pubyear{2015}
\volume{316}  
\jname{Formation, evolution and survival of massive star clusters}
\editors{C. Charbonnel and A. Nota eds.}

\begin{document}

\maketitle

\begin{abstract}
How starburst clusters form out of molecular clouds is still an open
question. In this article, I highlight some of the key constraints
in this regard, that one can get from the dynamical evolutionary properties
of dense stellar systems. I particularly focus on secular expansion of
massive star clusters and hierarchical merging of sub-clusters,
and discuss their implications \emph{vis-\'a-vis}
the observed properties of young massive clusters. The analysis suggests
that residual gas expulsion is necessary for shaping these clusters
as we see them today, irrespective of their monolithic or hierarchical
mode of formation.  
\end{abstract}

\firstsection 
\section{Introduction: birth conditions of young massive clusters}\label{intro}

Star clusters are found in our Milky Way and in all external galaxies with increasing detail.
However, how they form in the first place is still one of the most important challenges of
the cosmos. A key
question of wide debate is how the exposed clusters' parsec-scale,
centrally-pronounced, near-spherical shape, observed at all ages $\gtrsim 1$ Myr,
can be explained. This is in direct contrast with the irregular and much more extended
(10s of parsecs) structure of molecular clouds,
where stars form via gravitational fragmentation of dense substructures. An apparent lack
of an age spread among the members of the youngest star clusters (see, \eg, \citealt{bsv2013})
indicates that these stars form in a ``burst'' over a short period of time. This, in turn, implies that
short-timescale dynamical processes, \eg, violent relaxation \citep{spitz87},
immediately or simultaneously follow the
formation of the proto-stars, which shape the newly born cluster.

The unprecedented spatial and spectral resolution in IR and sub-mm wavelengths with
\emph{Herschel} and \emph{ALMA} has revealed intricate filamentary networks inside
dense regions of molecular clouds (see \citealt{andr2013} for a review). Such
observations generally infer that the (projected) widths of the individual filaments and their
junctions are as compact as 0.1-0.3 pc.
Both theoretical and observational studies suggest that groups of proto-stars preferentially form within
these filaments and at their junctions \citep{schn2012,andr2013}.

In \citet{sb2015a}, the role of violent relaxation in shaping a young cluster is
studied. It is demonstrated that in order to have a single cluster in dynamical equilibrium right from
Myr-age, as for massive starburst clusters, the stellar system involved in the mass assembly
process must be ``near-monolithic''. This implies the formation of either a single (monolithic) proto-cluster
within a dense molecular clump
or of several sub-clusters that merge in $\lesssim1$ Myr from parsec-scale separation.

In this study, useful constraints are obtained for assembling massive clusters given their birth conditions,
as summarized above, and the physics of dynamical evolution. Especially, I shall shed light on
one of the most widely debated question regarding massive cluster formation, namely, the relevance of primordial
gas blow-out by stellar feedback. Hereafter, the widely accepted definition of
young massive clusters (YMCs) is adopted, \ie, star clusters which have a present-day photometric mass exceeding
$\mphot\gtrsim10^4\Ms$ and are younger than $t\lesssim100$ Myr by stellar age. The youngest
subset of them, of $\lesssim4$ Myr age, are commonly referred to as `starburst'
clusters. These limits are not
robust but generally serve as representative values.

\begin{figure}[!t]
\centering
\includegraphics[width=10.0 cm,angle=0]{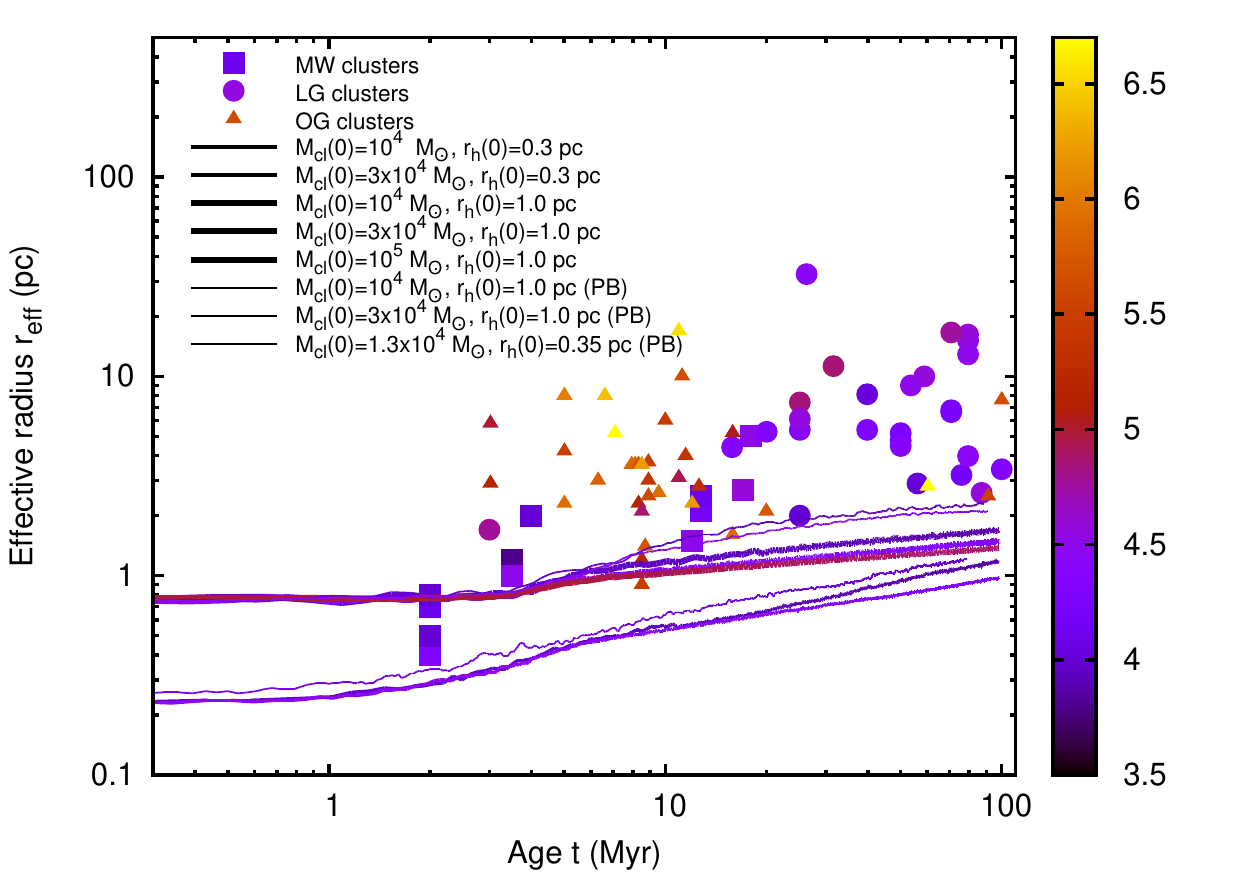}
\caption{Effective radius, $\reff$, vs. age, $t$, plot
(data from \citealt{pz2010}) for observed young
massive clusters (YMCs) in the Milky Way (MW), the Local Group (LG) and external galaxies (OG)
which are distinguished by different filled symbols. The symbols are colour-coded according
to the clusters' respective photometric mass, $\logten(\mphot/\Ms)$.
Overlaid are the computed curves for the
evolution of projected half-mass radius (or effective radius), $\reff(t)$, for model star
clusters, without residual gas expulsion (`PB'$\Rightarrow$ primordial binaries included),
which are also colour-coded by their instantaneous bound mass.}
\label{fig:reffevol_noexp}
\end{figure}

\section{Secular expansion of star clusters}\label{purexp}

A possible way in which a star cluster can appear within a molecular-gas filament is through a
localized but intense star formation at a privileged location in the filament, say, at a junction of
multiple filaments. Such a region is prone to reach a high SFE due to lateral contraction of the filaments
and ample gas supply through them. Any such stellar cluster (or sub-cluster)
should adapt to the compact cross-section of the filament (see Sec.~\ref{intro}).
While formation of proto-stars outside the filaments, \ie, in rarer gas,
cannot be ruled out, both observations (\eg, \citealt{schn2012,andr2013})
and hydrodynamic calculations (\eg, \citealt{bate2004,giri2011})
suggest that the majority of the proto-stars must form within the compact
dimensions of the dense gas filaments. Now, a cluster with $\reff\lesssim0.3$ pc is
way more compact than the presently observed YMCs \citep{pz2010},
\cf Fig.~\ref{fig:reffevol_noexp}. The key question here is whether such a compact star cluster
can expand by its own, through its secular evolution, to attain the presently observed sizes.

To reach the observed sizes, such a compact cluster must expand by a factor of $\gtrsim10$.
For a young, massive cluster, several effects contribute to its secular expansion.
In the earliest stage (until $t\approx4.5$ Myr), mass loss due to stellar
evolution is the primary driver of the expansion. The most massive stars remain
segregated (either primordially or dynamically) at the cluster's center, causing centrally localized
mass depletion due to the massive (O-type) stars' strong winds. When these stars 
undergo supernove, the central mass loss becomes even more severe, causing a higher rate
of cluster expansion. This stellar mass loss dominated expansion continues for $\approx50$ Myr.
After the stellar mass loss phase, the cluster expansion continues to be driven by
dynamical heating due to the centrally segregated black holes (hereafter BHs)
\citep{macetl2008,sb2010,morscher2013}.
Finally, young star clusters are observed to contain a substantial fraction ($\gtrsim70$\%) of 
tight primordial binaries of massive main-sequence stars \citep{sev2011}.
These primordial
binaries also inject energy to the cluster and contribute to its expansion
through binary-single and binary-binary encounters and
the associated binary heating and ejections of massive
stars \citep{hh2003}. 

To compute the secular evolution of model massive clusters in a realistic manner, the direct
N-body code {\tt NBODY7} \citep{aseth2012} is used.
In addition to computing the individual stars' orbits using the highly accurate
fourth-order Hermite scheme and dealing with the diverging gravitational forces, \eg, during close
encounters and in hard binaries through regularizations, {\tt NBODY7} includes
stellar and binary evolution recipes. The details of the computed models are
given in Table.~\ref{tab:complist_nogas}.
Some of the models contain a primordial binary population,
to assess the role of the latter in expanding a young cluster. In these models,
a $f_b(0)=100$\% primordial binary fraction that follow the
``birth orbital period distribution'' \citep{pk1995b} is used.
For massive stars of $m>5\Ms$, a much tighter period distribution,
given by a (bi-modal) \"Opik law (uniform distribution in $\logten P$) for
$0.3 < \logten(P/{\rm day}) < 3.5$ \citep{sev2011}, is used.

The solid curves in Fig.~\ref{fig:reffevol_noexp} shows the computed evolution of
effective radius, $\reff(t)$, for the models in Table~\ref{tab:complist_nogas}.
Here, the instantaneous $\reff(t)$
is obtained by taking the mean of the projected half-mass radii (50\% Lagrange
radius integrated over a plane) over three mutually perpendicular planes passing
through the cluster's density center. As seen, starting from sizes similar
to that of the filamentary substructures in molecular clouds (see Sec.~\ref{intro}),
it is practically impossible to attain
the observed sizes of YMCs and associations in 100 Myr.
Notably, realistic conditions are
used in these models including stellar mass loss,
retention of $\approx50$\% of the BHs and NSs formed via
supernovae and a realistic population of tight massive primordial binaries.  
A few test models also start with more extended size, $\rh(0)\approx1.0$ pc,
which can nearly reach the sizes of the most compact observed YMCs, but still are much more
compact than most YMCs. As elaborated in \citet{sb2015c}, theoretical uncertainties in
stellar and binary evolution and core-collapse supernova, would modify the
$\reff(t)$ curve to some extent, but are unlikely to alter the above overall
inference. This necessitates additional expansion mechanisms
for a newly assembled compact cluster to reach the present-day observed sizes.

\section{Non-secular evolution of star clusters: primordial gas expulsion}\label{gasexp}

The calculations in Sec.~\ref{purexp} do not consider any primordial gas present
initially in
the proto-cluster, \ie, the cluster is taken to be formed with 100\% local SFE.
A high local SFE has been
claimed by several hydrodynamics-based studies (\eg, \citealt{bate2004,dale2015}).
Clearly, as shown in Sec.~\ref{purexp}, if a cluster is hatched along a
molecular-gas filament or at a filament junction with effectively 100\% SFE,
it well falls short of observed sizes of YMCs. On the other hand,
observations of molecular clouds and embedded clusters suggest that in regions of
high star-formation activity, the local SFE
typically varies between a few percent to $\approx30$\%. This is furthermore
supported by high-resolution radiation magneto-hydrodynamic (RMHD) simulations of proto-star
formation \citep{mnm2012,bate2013}. Although these compute-expensive simulations are limited to the
spatial scale of a single proto-star, they suggest a maximum feedback-limited $\approx30$\% SFE
for proto-star formation; see \citet{sb2015b} for more discussion.
This implies that the SFE over a gas clump forming a population of proto-stars
is also $\lesssim30$\%. Hence, it would be realistic to consider a proto-cluster (a pre-cluster
cloud core) that is initially embedded in primordial gas,
the gas being subsequently cleared from the system by
the stellar feedbacks (\eg, radiation, outflows). 

\begin{figure}[!t]
\centering
\includegraphics[width=10.0 cm,angle=0]{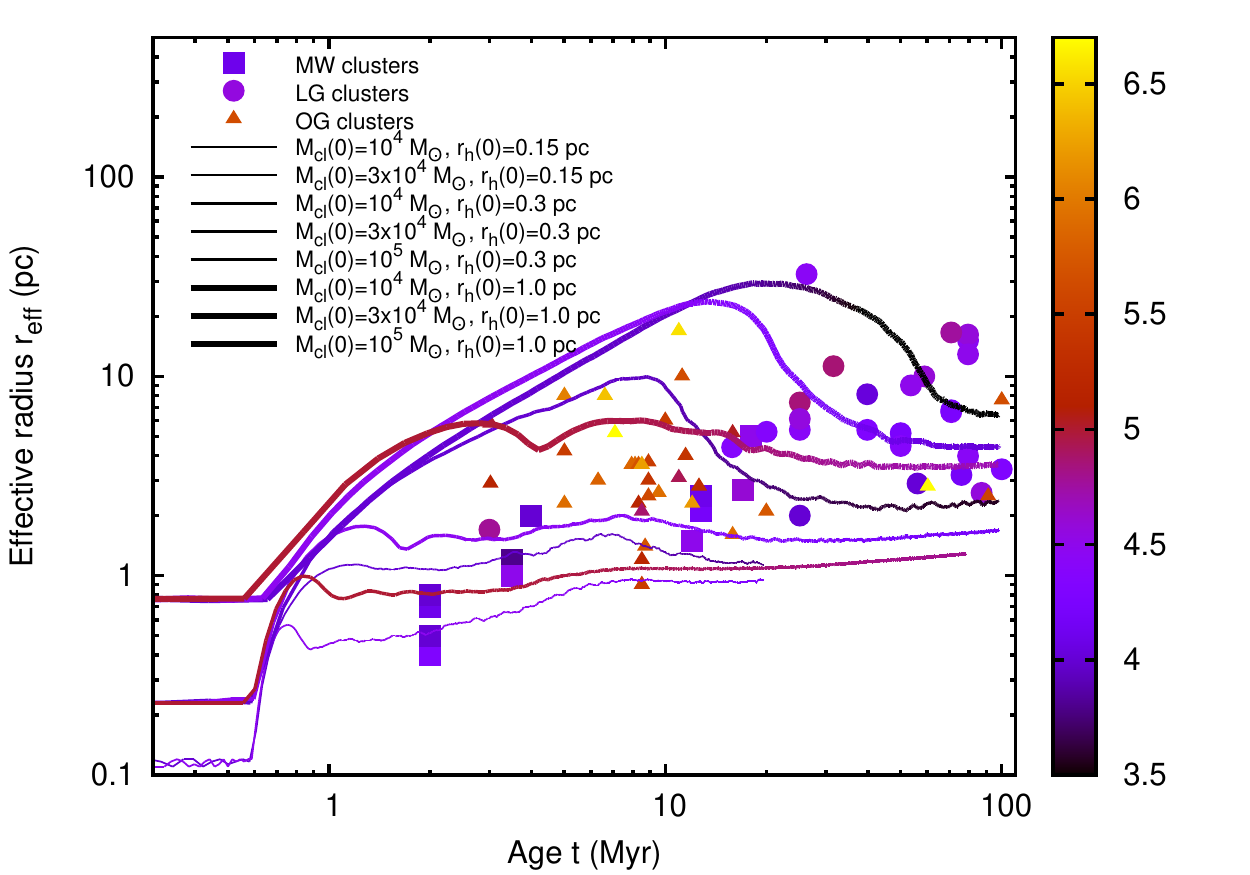}
\caption{The data points and their colour-code are
identical to Fig.~\ref{fig:reffevol_noexp}. The curves show the computed evolution of 
the effective radius, $\reff(t)$,
including a gas dispersal phase with star formation efficiency $\epsilon\approx33$\%.
These curves imply that even if YMCs evolve from filament-like
compact sizes, such substantial (and explosive) gas dispersal will expand them to their
present observed sizes in the Milky way and in the Local Group.
However, to attain the sizes of the most extended Local Group YMCs,
one needs to evolve from $\rh(0)\gtrsim1$ pc half-mass radii,
unless such objects are low-mass cluster complexes.}
\label{fig:reffevol_fastexp}
\end{figure}

Fig.~\ref{fig:reffevol_fastexp} shows the $\reff$ evolutions of computed model clusters
with $\approx30$\% SFE which are
compared to the observed $\reff$ values of YMCs and associations (upper and lower panels respectively).
Here, the `gas' is simply a time-varying background potential mimicking the expulsion of
ionized Hydrogen with sound speed, as implemented in previous studies (\eg, \citealt{pketl2001,sb2015a}).
With such an `explosive' (\ie, in timescale comparable to the dynamical time
of the cluster) gas expulsion of $\approx70$\% by mass, the filament-like compact clusters  
can expand to reach the observed sizes of most of the MW YMCs
(\cf Fig.~\ref{fig:reffevol_fastexp}).
However, most of the LG and OG YMCs are still a few factors
larger in size than the computed models with compact initial conditions
($\rh(0)\lesssim0.3$ pc). It seems likely that these extended YMCs are
ensembles of closely-located YMCs
($\sim10$s of them), forming low-mass cluster complexes, thus being
younger, low-mass versions of ``faint-fuzzy''
objects \citep{bru2009}; see \citet{sb2015c} for details.

\section{Cluster formation through hierarchical mergers}\label{himrg}

So far, monolithic or in-situ formation of star clusters are considered. 
The substructured and filamentary conditions in molecular clouds
(see Sec.~\ref{intro}) make it plausible that YMCs may also arise due to sequential
mergers of less massive sub-clusters, which
fall in the potential well of the molecular cloud (or clump).
In particular, if the sub-clusters are `decoupled' (\eg, as in \citealt{dale2015}) from
the gas, then the assembly process is independent and proceeds in a timescale longer
than the free-fall time of the gas.

If a cluster has to form and evolve from a young age (a few Myr
like the youngest Galactic YMCs), the sub-clusters
must fall in from sufficiently close separation so that they can merge
early enough. This is demonstrated in \citet{sb2015a} for the case of the
$\approx1$ Myr old NGC 3603 young cluster (photometric mass $\gtrsim10^4\Ms$),
for which the sub-clusters must
merge from $\lesssim2$ pc. This implies that despite star formation is
often found to occur over $\gtrsim10$ pc regions, only a part of the newly formed stellar
structure (typically of pc scale) can actually comprise a YMC.
There is mounting observational
evidence of young stellar and/or proto-stellar sub-clusters  
packed within pc-scale regions (\eg, \citealt{massi2014,rz2015}).

Moreover, as shown in \citet{sb2015a},
the ``prompt'' merger of several gas-filament-like compact sub-clusters produces a
similarly compact cluster. Therefore, according to the results in Sec.~\ref{purexp},
the observed YMC sizes would be unreachable for the newly assembled cluster
via their inherent expansion alone. This also holds for the individual sub-clusters. 
Therefore, as demonstrated in \citet{sb2015a}, a post-merger explosive gas expulsion is
instrumental in yielding YMCs that are like what we observe.   
If, on the other hand, the sub-clusters are brewed sufficiently apart that
gas blow out happens in them separately prior to their merger, then
the likely outcome would be a highly diffuse, massive stellar association
with substructures, and it may,
as a whole, be super-virial or sub-virial, \eg, Cyg-OB2 (also see \citealt{sb2015b}).

\section{Concluding remarks}\label{conclude}

The above study shows that if star
clusters preferably appear within the overdense filaments of molecular clouds,
adapting to their typical dimensions of 0.1-0.3 pc, then their self-driven
expansion is generally insufficient for reaching the dimensions of observed YMCs.
This is true if either the cluster forms in situ or via mergers of 
(closely located) sub-clusters, and holds irrespective of the newborn cluster's mass
(\cf Fig.~\ref{fig:reffevol_noexp}). Having run out of the other possibilities,
feedback-driven rapid expulsion of residual gas from the proto-cluster
(see Sec.~\ref{gasexp}) seems
to be indispensable to reach the observed sizes of YMCs. In this line
of argument, an important concern is how YMCs gather
$\gtrsim10^4\Ms$ within a few Myr, irrespective
of any formation channel. Such amount of mass reservoir, over pc scale,
is not immediately apparent from typical Galactic star-forming clouds. 
At the same time, the assembly phase of YMCs have to be short-lived ($\lesssim 1$ Myr),
as discussed above, implying that it would be rare to catch an assembling
starburst cluster.

The chances of finding a YMC in its assembling phase would be higher
in starburst galaxies, where a much larger number of massive clusters are triggered
compared to a Milky Way-like galaxy. Indeed,
recent ALMA observations of the Antennae galaxy indicate still-forming, deeply-embedded
stellar systems of total mass exceeding $10^7\Ms$, which are either monolithic or distributed
over a few pc \citep{john2015a,john2015b}.
In order to better understand the birth conditions of starburst clusters,
improved and more exhaustive observations of starburst
galaxies are necessary.

\begin{longtable}{rccccccc}
\caption{\label{tab:complist_nogas} Initial conditions for the computed model clusters
without a gas expulsion phase. The initial configurations
are Plummer profiles with total mass, $\mcl(0)$, and half-mass radius, $\rh(0)$.
The corresponding values of the virial velocity dispersion, $V_\ast(0)$,
the crossing time, $\tcr(0)$,
and the two-body relaxation time at half-mass radius, $\tauh(0)$, are given.
The initial clusters are in circular orbits at
$R_G\approx8$ kpc Galactocentric distance.}\\
\hline\hline
$\mcl(0)/\Ms$ & $\rh(0)/{\rm pc}$ & Primordial binaries  & Mass segregation & $V_\ast(0)/\kmps$ & $\tcr(0)/{\rm Myr}$ & $\tauh(0)/{\rm Myr}$\\  
\hline
\endfirsthead
\caption{continued.}\\
$\mcl(0)/\Ms$ & $\rh(0)/{\rm pc}$ & Primordial binaries  & Mass segregation & $V_\ast(0)/\kmps$ & $\tcr(0)/{\rm Myr}$ & $\tauh(0)/{\rm Myr}$\\  
\hline
\endhead
\endfoot
$10^4$ & 0.3 & no  & no & 10.5 & 0.028 & 4.65 \\ 
$3\times10^4$ & 0.3 & no  & no & 18.1 & 0.017 & 8.05 \\ 
$10^4$ & 1.0 & no  & no & 5.8 & 0.172 & 28.30 \\ 
$3\times10^4$ & 1.0 & no  & no & 10.0 & 0.100 & 49.02 \\ 
$10^5$ & 1.0 & no  & no & 18.2 & 0.055 & 89.49 \\ 
$1.3\times10^4$ & 0.35 & yes  & yes & 11.7 & 0.030 & 6.68 \\ 
$10^4$ & 1.0 & yes  & yes & 6.0 & 0.167 & 28.30 \\ 
$3\times10^4$ & 1.0 & yes  & yes & 10.4 & 0.096 & 49.02 \\ 
\hline\hline
\end{longtable}

\end{document}